\documentstyle[12pt,titlepage]{article}
\def\bib{\bibitem}
\def\be{\begin{equation}}
\def\ee{\end{equation}}
\def\beqar{\begin{eqnarray}}
\def\eeqar{\end{eqnarray}}
\def\barr{\begin{array}}
\def\earr{\end{array}}

\def\and{\qquad {\rm and } \qquad}

\def\p{\partial}
\def\f{F^a_{\mu \nu}}
\def\fn{F^{a \mu \nu}}
\def\g{G^a_{\mu \nu}}
\def\gn{G^{a \mu \nu}}
\def\gb{\bar{G}^a_{\mu \nu}}
\def\gbn{\bar{G}^{a \mu \nu}}
\def\gt{\widetilde{G}^a_{\mu \nu}}

\def\G{\bar{G}^{ab}_{\mu \nu}}
\def\gg{\frac{ \bar{G}_{\mu \sigma}\bar{G}^\sigma_{\; \nu}}
             {e^2 E^2}}
\def\gbe{\frac{i \bar{G}_{\mu \nu}}{eE}}

\def\pb{\bar{\psi}}
\def\min{\int \:d^4 x\:}
\def\path{\int {\cal D}A_\mu {\cal D}\pb {\cal D}\psi}

\def\slp{\partial \hspace{-1.5ex}/}
\def\sla{A \hspace{-1.5ex}/\:}
\def\sld{D \hspace{-1ex}/\:}
\def\slk{k \hspace{-1ex}/}
\def\slq{q \hspace{-1ex}/\:}
\def\lap{\rule{0.1mm}{1.5ex}\rule{0.7em}{0.1mm}\hspace{-0.72em}
         \rule[1.5ex]{0.7em}{0.1mm}
         \rule{0.1mm}{1.5ex}\:}
\begin{document}
\thispagestyle{empty}
\begin{flushright}
TIFR/TH/92--17  \\
IMSc/92--14
\end{flushright}
\vspace{5ex}

\centerline{\Large \bf Confinement In A Chromoelectric Vacuum}
\bigskip
\begin{center}
{\bf Rahul Basu}\footnote{E. mail address:
                           {\bf rahul@imsc.ernet.in}
                         }\\
{\it Institute of Mathematical Sciences\\
Madras -- 600 113, India}\\[5ex]
{\bf Debajyoti Choudhury}\footnote{E.mail address:
                                    {\bf debchou@tifrvax.bitnet}
                                  }\\
          {\it Theoretical Physics Group,
           Tata Institute of Fundamental Research \\
           Homi Bhabha Road, Bombay -- 400 005, India.
          }                         \\

\vspace{.8in}
{\bf Abstract}\\
\end{center}

\begin{quotation}
A chromoelectric vacuum that confines both gluon and quark
degrees of freedom (in the sense that they do not exist as
asymptotic states) is constructed. However some degrees of
freedom still exist as asymptotic states thereby allowing
colour singlets to propagate.
\end{quotation}
\vspace{.5in}

\newpage

The vacuum structure of QCD is a much studied subject. Many
attempts have been made in the 70's and early 80's to understand
the nature of the QCD vacuum. Various studies done many years
ago have shown conclusively that the perturbative vacuum is not
the true vacuum of QCD. A partial list of such attempts is given
in \cite{hist}. However, as is well known, there have been
numerous problems associated with many of these attempts. A
brief discussion of these problems is given in \cite{abp}.
Typically, these problems have been those of stability of the
vacuum (presence of a tachyon), presence of zero modes (infrared
divergences) and the like.

In an earlier publication \cite{abp} we had tried to address
some of these issues and had attempted to construct a
perturbation theory based on a different vacuum. While we do not
claim that this was the ``true'' vacuum of QCD, it was
nonetheless clear that the (chromoelectric) vacuum we had
constructed had many of the desirable features that we expect
the ``true'' vacuum of QCD to have. For one, there were no
tachyons thereby giving a stable vacuum where the effective
potential had no imaginary part unlike the Saviddy vacuum. Also,
some of the colour degrees of freedom were absent as physical
excitations and to first order there were no massless modes
thereby ameliorating the IR problem considerably. However, the
drawback was that the study in
\cite{abp} was based on pure $SU(3)$ gauge theory with no fermions.

In this paper, we have extended our formalism to include
fermions (quarks). As in \cite{abp}, we have
incorporated the effect of the quartic vertex through the
introduction of an auxilliary field which then renders the
theory cubic. A classical background for this auxilliary field
then leads to modifications in the perturbative vacuum as well
as the tree--level two--point function for the gluons. The
quarks interact with the background only at the
one--loop order to get confined in turn. For this
purpose, as before, we are constrained to work in Minkowski
space and hence use Minkowski functional integral formalism.

For completeness, we give below the formalism that we have used
in \cite{abp} but with an additional term for massive fermions.
For a gauge theory with massive fermions  the partition function
is
\be \displaystyle
  Z = \path e^{\min i\left[ \frac{1}{4g^2} \f \fn +  \pb
                           (i \sld -m)  \psi \right]},
\ee
where
\be
\barr{rcl}
  D_\mu \psi &=& (\p_\mu +  A_\mu^a T^a) \psi \\
  \f &=& \p_\mu A_\nu^a - \p_\nu A_\mu^a +
            f^{abc}A_\mu^b A_\nu^c.
\earr
\ee
As stated above we are working with a Minkowski space (signature
$-2$) field theory.  This, as we shall see later, is important.

On introducing an auxiliary field $\g$ and using
\be
1 = \int {\cal D}G_{\mu\nu}\exp \left[-\frac{i}{4}\int d^4 x
                                    (g\g-{1\over g}\f)^2 \right],
\ee
the gauge--fixed partition function can be written, upto a
constant term, as
\be \displaystyle
  Z = \path{\cal D}G_{\mu\nu}{\cal D}c {\cal D}\bar{c}
                       \:e^{\min i \left[ \frac{-g^2}{4} \g \gn +
                          \frac{1}{2} \g \fn +  \pb (i \sld -m)
                      \psi \right] + S_{GF} + S_{FPG}}
\ee
where $c,\bar{c}$ are the ghost fields and $S_{GF}$ and
$S_{FPG}$ the gauge fixing and the Faddeev--Popov ghost terms
respectively. The introduction of the auxilliary field has the
benefit of making the theory look a cubic one, with the quartic
interaction terms being absorbed in the $\g \gn$ term.

We now choose to evaluate the partition function in the presence
of a background gauge field and background auxiliary field.
\be
\barr{rcl}
 \g &=& \gt + \gb.  \\
  A_\mu^a &=& \bar A_\mu^a +\tilde A_\mu^a.
\earr
\ee

The interpretation of this split is quite simple : $\gb$
represents the zero--momentum mode in the Fourier expansion
while $\gt$ gives the orthogonal non--zero momentum components.
The same is true of $A_\mu^a$.  Now $\gb$ being a group valued
antisymmetric field, in $3 + 1$ dimensions, the most general
expression for it is then
\cite{abp}
\be
\gb = e(n_\mu E_\nu^a- n_\nu E_\mu^a) +
      h \epsilon_{\mu \nu \sigma \lambda}n^\lambda H^{a \sigma},
\ee
where $e,h$ are scalar constants of mass dimension 2, $n_\mu$ is
an arbitrary time--like vector normalized to $n^2 = 1$ and $E^a$
and $H^a$  are group valued vectors orthogonal to $n$.

Though one can proceed with this general formulation, it has
been shown
\cite{abp} that for any non--zero $h$ tachyonic modes are
present in the theory. Hence we make the choice of $h=0$. The
relevant quantities are then $e$ and $E_\mu^a$ which are then
unique upto group or space rotations. For simplicity,
we take the gauge fixing term to be
\be
{\cal L}_{GF} = \frac{1}{2g^2}(D_\mu^{ab} A^{\mu b})^2.
\ee
We again make the ansatz $\bar A_\mu^a =0$ to maintain manifest
translation invariance.

The path integral measure can then be split into ${\cal D}\gt$
and $d\gb$ to yield (here the tilde on $\g$ has been dropped for
convenience)
\be \displaystyle
  Z = \int d \gb e^{\frac{-ig^2}{4}\min \gb \gbn }
      \path{\cal D}G_{\mu\nu}{\cal D}c {\cal D}\bar{c}
                       \:e^{i S_{\rm free} +i S_{\rm int} }
\ee
with
\be
\barr{rcl}
{\cal L}_{\rm free} &=& \displaystyle \frac{-g^2}{4}\g \gn
                 - \frac{1}{2 g^2} A^{a\mu}
                 \left(\lap \delta^{ab}
                 \eta_{\mu \nu} +g^2 \G \right)A^{b\nu}
                + \pb (i\slp -m) \psi + \bar{c}^a \lap c^a     \\
{\cal L}_{\rm int} &=& \displaystyle \frac{1}{2} f^{abc}
                     \left\{ \g + \frac{1}{g^2}
                             (\p_\mu A_\nu^a - \p_\nu A_\mu^a)
                             \right\}
                      A^{b\mu} A^{c\nu}
               + f^{abc} \bar{c}^a \p^\mu (A^b_\mu  c^c)
               + i \pb \sla \psi,
\earr
\ee
where
\be
  \G = f^{acb} \bar{G}^c_{\mu \nu}.
\ee
The Feynman rules can then easily be read off \cite{abp} from
the above expressions after doing the $d \gb$ integral by the
stationary phase approximation and minimizing the effective
potential with respect to $e$. While the gauge--fermion vertex
as well as all pieces involving the ghost fields are exactly the
same as in usual QCD, the gauge field propagator changes
drastically on account of the background field. Denoting
\be
      E_\mu^{ab} = f^{acb}E_\mu^c,
\ee
the two point function in momentum space now reads
\be
\barr{rcl}
D^{ab}_{\mu \nu}(p^2) \equiv \langle A_\mu^a A_\nu^b \rangle_0
  &=& \displaystyle
    \frac{-ig^2}{p^2} \left[ (\eta_{\mu \nu} - n_\mu n_\nu )
       \delta^{ab}
       -\left( E_\mu \frac{1}{E^2} E_\nu \right)^{ab}\right] \\
&&\quad \displaystyle
   -i \left[ \frac{g^2}{p^4 + g^4 e^2 E^2} \right]^{ac}
       \left[p^2 n_\mu n_\nu + E_\mu \frac{p^2}{E^2} E_\nu
    +e g^2 (n_\mu E_\nu - n_\nu E_\mu )
                          \right]^{cb}
\earr
\ee
and similarly for $\langle \g G_{\lambda \sigma}^b \rangle$ and
$\langle \g A_\lambda^b \rangle$ as given in \cite{abp}. The
denominator in the second piece obviously leads to propagator
poles off the real axis. In fact, a careful study shows that in
$d$ dimensions, for a $SU(N)$ theory one has $(N^2 -1) (d-2)$
massless poles and the rest are complex conjugate pairs.  The
latter modes then do not propagate in position space and lead to
confinement. The contribution of the propagating modes, on the
other hand, is cancelled by the ghost contributions. Of course,
one might raise the objection that the formulation itself is
Lorentz non--covariant (on account of the presence of the
arbitrary vector $n_\mu$) and the results are mere artifacts of
this treatment. That this is not so can clearly be seen by
expressing the propagators in terms of $\gb$, {\it e.g.}
\be
D^{ab}_{\mu \nu}(p^2) =
      -ig^2 (p^2 \eta_{\mu \nu}\delta^{ab} -g^2 \G )^{-1}.
           \label{cov prop}
\ee
This then assures completely Lorentz--invariant dispersion
relations.

To have a more definite idea of the propagator poles, we then
make a particular choice namely
\[
n_\mu = (1,0,0,0).
\]
The only non--vanishing components of $E_\mu$ are then the space
components and in four dimensions these can always be chosen to
satisfy the $SU(2)$ algebra namely
\be
[E_i, E_j] = \epsilon_{ijk} E_k \and [E^2, E_\mu] = 0.
\ee
For definiteness, we make the ansatz
\[
(E_1)^{ab} = 2 f^{2ab}, \quad
(E_2)^{ab} = 2 f^{4ab}, \quad
(E_3)^{ab} = 2 f^{6ab}.
\]
{}From eqn.(\ref{cov prop}), we see that the poles of the
propagator are given by the eigenvalues of $\G$. For a $SU(3)$
theory in 4 dimensions, these are at
\be
p^2 = \left\{
     \matrix{0\: (16)\cr
             \pm i\sqrt{6}e g^2 \: (5 \;{\rm pairs})\cr
             \pm i\sqrt{2}e g^2 \: (3 \;{\rm pairs})\cr} \right.
\ee
where the degeneracy is shown in parentheses. Though we have
evidence of propagator poles off the real axis, before we make
any statement about confinement we must be sure that the
background configuration we have chosen actually has a energy
lower than that of the perturbative vacuum. Such a comparison is
to be made for the effective potential. Remembering that the
interaction terms in the Lagrangian do not contribute to the
tree--level effective action, the latter is trivially evaluated
to yield
\be
\barr{rcl}
  \Gamma^{\rm eff}_0 (0,0,0) &=& \displaystyle
        -\frac{g^2}{4} \gb \gbn
        + {\rm Tr}\: \frac{1}{2i} \int \frac{d^d q}{ (2 \pi)^d}
          \ln \left[ -i (q^2 \eta_{\mu \nu}\delta^{ab} -g^2 \G )
              \right]
                           \\
 && \displaystyle
        - {\rm Tr}\: \frac{1}{i} \int \frac{d^d q}{ (2 \pi)^d}
            \ln (i q^2)
        -  \int \frac{d^d q}{ (2 \pi)^d}
           \ln (\slq -m) .
\earr
\ee

Concentrating on the gauge part, we see that the renormalized
effective  potential for our choice of the $SU(N)$ background reads
\be
  \Gamma^{\rm eff}_{0\:{\rm ren} } =
          \frac{g^2 e^2}{2 N} {\rm Tr} \: E^2
         + \frac{g^4 e^2}{16 \pi^2} {\rm Tr} \:
            \left[ E^2 \ln \frac {e^2 E^2}{\mu^4} \right] + \cdots
\ee
where $\mu$ is the renormalisation scale and $\cdots$ represents
the fermionic contribution. Minimising $  \Gamma^{\rm
eff}_{0\:{\rm ren} } $ with respect to $e^2$ we get
\be
 \Gamma_{{\rm min} } =
          - \frac{g^4}{16 \pi^2} e^2_{\rm min} {\rm Tr} \:
            \left[ E^2 ( 1 + \ln E^2 ) \right]
            + \cdots
\ee
where
\be
   \ln \frac{ e^2_{\rm min} } { \mu^4 }
         = -\left[ 1 + \frac{8 \pi^2}{N g^2} +
                  \frac { {\rm Tr} \left( E^2 \ln E^2 \right) }
                        { {\rm Tr} \left( E^2 \right) }
            \right]
\ee
This explicitly shows that the configuration we have chosen
indeed leads to vacuum that is lower than the perturbative one.

We now turn to the question of quark propagation in the theory.
Though our choice for the background does not seem to involve
the  fermions, the latter would be affected on account of
radiative corrections. As there is no $G \pb \psi$ coupling,
there is obviously no correction at the tree level. In fact, the
only correction to the fermionic propagator at the one loop
level comes from the usual graph, but with the modified gauge
propagator. That this is so becomes obvious once one realises
that $D_{\mu \nu}^{a b} (p^2)$ represents the \underline{full
propagator} at the tree--level, {\it i.e.}, it is the quadratic
operator in the effective action. Also the fact that there is no
$G \pb \psi$ coupling in the Lagrangian assures that the graph
being discussed is the only one contributing to the one--loop
order, provided of course one uses the full zero--loop gluon
propagator. A straightforward calculation can then be made, but
with extreme care as to Wick rotation. To this end, it is more
useful to reexpress the gluon propagator as
\be
D^{ab}_{\mu \nu}(p^2) =
       -i g^2 \sum_{j=1}^3 \left( \frac{1}{p^2 - M_j^2}
                           \right)^{ac}
                          L_{\mu \nu}^{cb}(j)
\ee
where
\be
\barr{rcl}
  M_1^2 & = & -i \epsilon \\
  M_2^2 & = & i e g^2 E \\
  M_3^2 & = & - i e g^2 E \\
  L_{\mu \nu}^{ab}(1)
         &=& \displaystyle
          \eta_{\mu \nu} \delta^{ab}
          -\left( n_\mu n_\nu + E_\mu \frac{1}{E^2} E_\nu
           \right)^{ab}
                              \\[1ex]
&=& \displaystyle
 \eta_{\mu \nu} \delta^{ab} + \left( \gg \right)^{ab}
\\[2ex]
  L_{\mu \nu}^{ab}(2) &=& \displaystyle
     \frac{1}{2} \left[ n_\mu n_\nu + E_\mu \frac{1}{E^2} E_\nu
                     -\frac{i}{E} (n_\mu E_\nu - n_\nu E_\mu )
                  \right]^{ab}
                               \\[1ex]
 &=& \displaystyle
    -\frac{1}{2} \left[\gbe + \gg \right]^{ab}
                               \\[2ex]
  L_{\mu \nu}^{ab}(3) &=& \displaystyle
     \frac{1}{2} \left[ n_\mu n_\nu + E_\mu \frac{1}{E^2} E_\nu
                       +\frac{i}{E} (n_\mu E_\nu - n_\nu E_\mu )
                \right]^{ab}
                               \\[1ex]
&=& \displaystyle \frac{1}{2} \left[\gbe - \gg \right]^{ab}      \\
\earr
\ee

The one--loop correction to the fermion propagator is then given
by
\be
\barr{rcl}
  i\Sigma_{\alpha \beta} &=& \displaystyle
              \mu^{4-d} \int \frac{d^d q}{(2 \pi)^d}
                      i (T^a)_{\alpha \gamma} \gamma^\mu
                   D^{ab}_{\mu \nu}(q^2) \frac{i}{\slk -\slq -m}
                             i (T^b)_{\gamma \beta} \gamma^\nu
                             \\
&=& \displaystyle  -g^2 (T^a T^b)_{\alpha \beta} \sum_{j = 1}^3
                 L_{\mu \nu}^{cb}(j) \gamma^\mu J^{ac}(j)
                 \gamma^\nu
\earr
\ee
where
\be
J^{ac}(j) = \mu^{4-d} \int \frac{d^d q}{(2 \pi)^d}
               \frac{\slk -\slq + m}{(k - q)^2 - m^2}
             \left( \frac{1}{q^2 - M^2_j} \right)^{ac}
\ee
and $\mu$ is the mass parameter introduced in dimensional
regularization.

The usual course of action at this point is to employ either the
Feynman or the Schwinger parametrization for the denominator of
the integrand followed by the Wick rotation. A naive application
of such methods in this context is however fraught with danger
as one is dealing with a matrix valued gauge propagator and what
is more one with poles off the real axis.  It is then that the
separation of the propagator into three pieces exhibits its
usefulness. The matrix $E$ being a diagonal one with all
eigenvalues positive, the Feynman parametrization can be
employed for each of the three pieces in a relatively
straightforward fashion.  Realizing that the Wick rotations for
the $j=2$ case  has to  be made in a sense opposite to that for
$j = 1$ and $j =3$, we can Euclideanize the integrals to have
\be
J^{ac}(j) = -i 
               h_j \mu^{4-d} \int_0^1 dx\:
                \int \frac{d^d q_E}{(2 \pi)^d}
                     \left[
                       \frac{x \slk_E + m}
                            { \left( q_E^2 + \xi_j^2 \right)^2}
                     \right]^{ac}
\ee
where
\be
\barr{rcl}
\xi_j^2 &=& x(1-x)k^2_E + x M_j^2 + (1-x) m^2
                \\
h_j &=& \left\{
               \barr{rl}
                   1 & j = 1, 3 \\
                  -1 & j = 2
                \earr
                         \right.
\earr
\ee
A straightforward computation then yields
\be
\barr{rcl}
\displaystyle 16 \pi^2 i h_j J^{ac}(j) &= &
              \displaystyle \int_0^1 dx\: (x \slk_E +m)
               \left[ \frac{2}{\varepsilon} - \gamma_E -
                     \ln \frac{\xi_j^2}{4 \pi \mu^2} \right]^{ac}
                     \\
& =&\displaystyle
               \left(\frac{1}{2} \slk_E + m \right)
                  \left( \frac{2}{\varepsilon} - \gamma_E
                        -\ln \frac{k_E^2}{4 \pi \mu^2}
                \right) \delta^{ac}
                       \\
 & &\displaystyle
       - \frac{\slk_E}{2}
              \left[  - (x_{j+}+ x_{j-})
                         + (1 - x_{j+}^2 ) \ln (1 -x_{j+})
              \right.
                       \\
 && \displaystyle \qquad \left.
                   + x_{j+}^2 \ln x_{j+}
                  + (1 - x_{j-}^2 ) \ln (1 -x_{j-})
                  + x_{j-}^2 \ln (-x_{j-})
                     \right]^{ac}
                         \\
 && \displaystyle - m
              \left[ - 2
                     + (1 - x_{j+} ) \ln (x_{j+}-1)
                     + x_{j+} \ln x_{j+}
                     \right.
                             \\
 &&  \displaystyle \qquad \qquad \left.
             + (1 - x_{j-} ) \ln (1 -x_{j-})
             + x_{j-} \ln (x_{j-})
                                         \right]^{ac}
\earr
\ee
where
\be
\barr{rcl}
    x_{j \pm} &=&\displaystyle
               \frac{1}{2 k_E^2}
               \left[ k_E^2 + M_j^2 - m^2 \pm
                     \sqrt{(k_E^2 + M_j^2 -m^2)^2 + 4 m^2 k_E^2}
             \;\right]  \\
    \gamma_E &=& {\rm Euler - Mascheroni \; constant} \\
    \varepsilon &=& 4 -d
\earr
\ee
As the above equations stand, they are exact but shed little
light. To make things more transparent we shall work under the
approximation
\be
   m^2 \ll k_E^2, \left| M_{2,3}^2 \right|
\ee
On reexpressing everything in Minkowski space, the expression for the
one--loop correction then simplifies to
\be
  i\Sigma_{\alpha \beta} (k)
= \displaystyle  -\frac{ig^2}{16 \pi^2} (T^a T^b)_{\alpha \beta}
                 \gamma^\mu
          \left( B_{\mu \nu}^{ab}
             - \frac{i}{2} C_{\mu \nu}^{ab}  \right) \gamma^\nu
\ee
with
\be
\barr{rcl}
B_{\mu \nu}^{ab}  &=& \displaystyle
        \left( \frac{2}{\varepsilon} - \gamma_E -
                 \ln \frac{- k^2}{4 \pi \mu^2} + 2 \right)
        \left( -\frac{1}{2}\slk + m \right)
        \left[ \eta_{\mu \nu} + \gbe
           + \gg
            \right]^{ab}             \\
C_{\mu \nu}^{ab}  &=& \displaystyle
   - \slk
            \left[ \gbe
                   \left\{ (1-a^2) \ln (1+a^2) + a^2 \ln a^2
                              + 2 a (\pi - 2 \tan^{-1} a) \right\}
            \right. \\
  &&\displaystyle         \qquad \left.
           + \gg
           \left\{ -a + \frac{\pi a^2}{2}
                      + (1 - a^2) \tan^{-1} a
                      - a \ln \frac{a^2}{1+ a^2} \right\}
                     \right]^{ab} \\
&&\displaystyle
+  m
            \left[ \gbe
                   \left\{ \ln (1+a^2) + a (\pi - 2 \tan^{-1} a)
                    \right\}
              + \gg \left\{2 \tan^{-1} a
                           - a \ln \frac{a^2}{1+ a^2}
                    \right\}
                     \right]^{ab} \\
\earr
\ee
where
\be
 a^{ab} \equiv - g^2 \frac{eE^{ab}}{k^2}
\ee

The matrix $C_{\mu \nu}^{ab}$ is real and symmetric under the
interchange of indices $(\mu \: a) \leftrightarrow (\nu \: b)$
and hence has real eigenvalues. Its presence in  $\Sigma_{\alpha
\beta}$ thus renders the latter's eigenvalues complex. The
corrected fermion propagator is given by
\be
 S_F(k) = \displaystyle \frac{i}{\slk - m - \Sigma(k)},
\ee
and hence has poles off the real axis at the one-loop level.  It
is obvious that the complex poles are essentially given by those
of the gluon propagator. This feature is to be expected as all
such behaviour in the theory owe their origin to a non--zero
value of the background gauge $\gb$. As in the gluonic case,
even here only some colour degrees of freedom get confined and
not all. This is as well, because this feature allows the color
singlet degrees of freedom to propagate.

Looking at the large $a$ limit of the massless theory, one sees
that the one--loop fermion propagator still goes as
\be
 S_F(k_E) \sim \frac{i}{\slk}
\ee
as is only to be expected,
for the choice of the classical background does not affect the
chiral properties of the thory in any way.

 An interpretation of the finite terms in the theory is somewhat
tricky.  While some of these just result in finite wavefunction
and mass corrections, the others have the appearance of dipole
moments. Such an identification though would be incorrect as
these do not represent true interaction terms but only the
effect of the background gauge.

In conclusion, we have shown that it is possible to construct a
chromoelectric vacuum for non-Abelian gauge theories that
confines some of the gluon and quark degrees of freedom in the
sense that they do not exist as asymptotic states. This vacuum
also considerably ameliorates the infra--red problems due to the
absence of massless gluon poles at least to first order. In
addition, since not all the degrees of freedom of the theory are
confined, colour singlets are still allowed to propagate.

One outstanding question in this analysis is that of
renormalisability.  As is clear from the expression for
$B_{\mu\nu}^{ab}$ in (31), the infinite terms are now functions
of the background field $\bar G_{\mu\nu}$ as expected. Hence,
even though terms linear in $\bar G_{\mu\nu}$ can be removed
through a global colour averaging of the $\bar G_{\mu\nu}$
field, terms of higher order in $\bar G_{\mu\nu}$ can still pose
a potential problem and therefore a careful examination of the
structure of the infinite terms, perhaps through a study of
Slavnov--Taylor type identities of this theory, is needed. The
presence of complex poles also necessitates a study of the
unitarity of the theory.  These issues are currently being
examined.

{\bf Acknowledgements} We would like to thank R. Anishetty and
R. Parthasarathy for many useful discussions. We also thank R.
Parthasarathy for a careful reading of the manuscript.

\end{document}